\documentclass[fleqn,twoside]{article}
\usepackage[headings]{espcrc2}

\usepackage{graphicx,times,psfig}       

\newcommand{\AmS}{{\protect\the\textfont2
  A\kern-.1667em\lower.5ex\hbox{M}\kern-.125emS}}

\newcommand{\degr}{^{\circ}}

\title{New knowledge of the Galactic magnetic fields}

\author{JinLin Han\address[]{National Astronomical Observatories, 
Chinese Academy of Sciences, Beijing 100012, China}}

\begin{document}

\begin{abstract}
The magnetic fields of our Milky Way galaxy are the main agent for cosmic
rays to transport. In the last decade, much new knowledge has been gained
from measurements of the Galactic magnetic fields. In the Galactic disk,
from the RMs of a large number of newly discovered pulsars, the large-scale
magnetic fields along the spiral arms have been delineated in a much larger
region than ever before, with alternating directions in the arm and interarm
regions. The toroidal fields in the Galactic halo were revealed to have
opposite directions below and above the Galactic plane, which is an
indication of an A0 mode dynamo operating in the halo. The strength of
large-scale fields obtained from pulsar RM data has been found to increase
exponentially towards the Galactic center. Compared to the steep Kolmogorov
spectrum of magnetic energy at small scales, the large-scale magnetic fields
show a shallow broken spatial magnetic energy spectrum.
\vspace{1pc}
\end{abstract}

\maketitle

\section{Introduction}           
\label{sect:intro}

Our home Galaxy, the Milky way, is very special in the universe. Supernova
explosions in our Galaxy are main accelerators for the middle (TeV) or
high (EeV) energy cosmic rays. On the other hand, our Galaxy is a large
deflector for cosmic rays because of the magnetic fields in interstellar
space.

The first idea about the Galactic magnetic fields was proposed by Fermi
\cite{fer49} when he suggested the origin of cosmic rays from interstellar
space and the acceleration by interstellar magnetic fields. Richtmyer \&
Teller found that magnetic field of $10^{-6}$G is needed to confine the
cosmic rays with energy $E \sim 10^{14}$~ev in a scale of $r \sim
3*10^{17}$~cm \cite{rt49}.  Though Alfv\'en \cite{alf49} insisted for the
solar origin of cosmic rays, he first estimated the strength of Galactic
fields amplified by the motion of interstellar medium should be a few $\mu$G,
which is correct, using the equipartition of magnetic field energy with
motion of gas in the form of $B^2/8\pi \sim \rho v^2/2$ and adopting
interstellar gas density $\rho \sim 10^{-24}$g~cm$^{-3}$ and typical gas
velocity of 10~km~s$^{-1}$. These were only very basic ideas on the extent
and strength of Galactic magnetic fields, which are quite right even
according to recent analysis \cite{bc90}. When the extremely high energy
cosmic rays ($>10^{18}$eV) were detected or even imagined, much advanced ideas
on the extragalactic magnetic fields were very first proposed,
e.g. \cite{bb72,bo99}. Nowadays, cosmic ray experts have modeled the
magnetic fields of, even outside, our Galaxy and try to simulate the
propagation and deflection of cosmic rays as well as their anisotropy,
e.g. \cite{tys06,yns04,ps03,tt02}. It seems that cosmic ray experts often have
many advanced ideas about magnetic fields, while observational measuring of
the magnetic fields is apparently rather difficult.

The Galactic magnetic field can be detected through observations of Zeeman
splitting of spectral lines, of polarized thermal emission from dust at mm,
sub-mm or infrared wavelengths, of optical starlight polarization due to
anisotropic scattering by magnetically-aligned dust grains, of radio
synchrotron emission, and of Faraday rotation of polarized radio sources
\cite{hw02}. The first two approaches have been used to detect respectively
the line-of-sight strength and the transverse orientation of magnetic fields
in molecular clouds \cite{ncr+03,hc05}. Starlight polarization can be used
to delineate the orientation of the transverse magnetic field in the
interstellar medium within 2 or 3 kpc of the Sun. Careful analysis of such
data shows that the local field is mainly parallel to the Galactic plane and
follows local spiral arms \cite{hei96}. Since we live near the edge of the
Galactic disk, we cannot have a face-on view of the global magnetic field
structure in our Galaxy through the polarized synchrotron emission, which
is possible for nearby spiral galaxies \cite{bbm+96}. Polarization observations
of synchrotron continuum radiation from the Galactic disk give the
transverse direction of the field in the emission region and some indication
of its strength. Large-angular-scale features are seen emerging from the
Galactic disk, for example, the North Polar Spur \cite{rfr+02}, and the
vertical filaments near the Galactic center \cite{dhr+98}. There are also
many small-angular-scale structures resulting from the diffuse polarized
emission at different distances which are modified by foreground Faraday
screens. Faraday rotation of linearly polarized radiation from pulsars and
extragalactic radio sources is a powerful probe of the diffuse magnetic
field in the Galaxy \cite{sk80,sf83,ls89,rk89,hq94,hmbb97,id99}. Faraday
rotation gives a measure of the line-of-sight component of the magnetic
field. Extragalactic sources have the advantage of large numbers but pulsars
have the advantage of being spread through the Galaxy at approximately known
distances, allowing direct three-dimensional mapping of the field. Pulsars
also give a very direct estimate of the strength of the field through
normalization by the dispersion measure (DM).

A diffuse magnetic field exists on all scales in our Galaxy. In fact, only
if the magnetic field $\vec{B}(x,y,z)$ in all positions in our Galaxy is
known, one can say that we have a complete picture of the Galactic magnetic
fields. Here, $x,y$ are in the Galactic plane, and $z$ is normal to that. In
practice, we only have ``partial'' measurements in some regions, so we never
get a complete picture. However, if we ``connect'' the available
measurements of different locations, then we may outline some basic features
of the Galactic magnetic fields. When one tries to look at the large-scale
field structure, the small-scale fields act as ``random'' fields,
``interfering'' with your measurements to an extent that depends on the
strength of random fields. The large-scale magnetic fields appear as a kind
of smooth background at small scales, and exist in a coherent manner at
different locations inside our Galaxy. To describe the Galactic magnetic
fields, we need to clarify following items:
\vspace{-1mm}
\begin{itemize}
\item {\bf Field structure}
  \begin{itemize}
      \item Disk field: local structure in the Solar vicinity (3 kpc)?
      \item Disk field: large scale structure and reversed directions
 in arm and interarm regions?
      \item Field structure in the Galactic halo?
      \item Field structure near the Galactic Center?
  \end{itemize}
\item {\bf Field strength $B$}
  \begin{itemize}
      \item Random field versus ordered field: $\langle \delta B \rangle^2$
      vs. $B^2$?
      \item Variation of field strength with the Galacto-radius
	($R=\sqrt{x^2+y^2}$), i.e. $B$ or $\delta B$ varies with $R$?
      \item Variation of field strength with the Galactic height ($z$): 
        $B$ or $\delta B$ varies with $z$?
      \item $B$ or $\delta B$: difference in arm and interarm regions?
  \end{itemize}
\item {\bf Fluctuations and scales}
  \begin{itemize}
      \item Spatial B-energy spectrum in large and small scales?
      \item Maximum field strength in the energy injection scale?
  \end{itemize}
\end{itemize}
\vspace{-1mm}

In this review, the magnetic fields in our Galaxy we talk about are the
fields in the diffuse interstellar medium, rather than the fields in
molecular clouds which are very extensively reviewed by Heiles \& Crutcher
\cite{hc05}.

\section{The Galactic magnetic fields: a decade ago}

\noindent{\bf A. Local disk field: there was some consensus.} 

Starlight polarization data are mostly limited to stars within 2 or 3 kpc
but gave the very first evidence for large-scale magnetic fields in our
Galaxy \cite{mf70}. It has been shown that the local field is parallel to
the Galactic plane and follows the local spiral arms \cite{am89}.  The
rotation measures (RMs) of small sample of local pulsars showed that the
local magnetic field going toward $l\sim90^o$ \cite{man74}, with a strength
about 2~$\mu$G. The field reversal near the Carina-Sagittarius arm was shown
by model-fitting to the pulsar RM data by Thomson \& Nelson \cite{tn80}. All
these have been confirmed later by much more pulsar RM data \cite{hq94,id99,hmq99}.

\vspace{1mm}
\noindent{\bf B. Which model for the large-scale disk field}

When available measurements are very limited, a good model is needed to
connect the measurements and give an idea of the basic features.
Simard-Normandin \& Kronberg \cite{sk80} and Sofue \& Fujimoto \cite{sf83}
showed that the large-scale magnetic fields in the Galactic disk with a
bisymmetric spiral (BSS) structure of a negative pitch angle and field
reversals at smaller Galacto-radii can fit the (average) RM distribution of
extragalactic radio sources along the Galactic longitudes better than the
concentric ring model and the axisymmetric spiral (ASS) field model. After
the RMs of 185 pulsars were published \cite{hl87}, the field reversals near
the Carina-Sagittarius arm and the Perseus arm were confirmed
\cite{ls89}. Rand \& Kulkarni \cite{rk89} failed to fit the BSS to the
pulsar RM data and emphasized the validity of the concentric ring model
(also \cite{rl94}). Vall\'ee argued \cite{val96} for an axisymmetric spiral
field model according to early RM data of extragalactic radio sources near
tangential directions of spiral arms. Han \& Qiao \cite{hq94}carefully
checked the model and data, and found that the BSS model is the best to fit
pulsar RM data. 

It seems to be not very clear which structure the Galactic magnetic 
fields have.

\vspace{1mm}
\noindent{\bf C. Fields in the Galactic halo or the thick disk}  

Evidence was found for a thick magneto-ionic disk with a scale height of
$\sim$1.2~kpc \cite{sk80,hq94}. A thin and thick radio disks were found from
modeling the radio structure of our Galaxy at 408~MHz \cite{bkb85}.
Large-scale polarized features in sky (see a comprehensive review by Reich 
\cite{rei06} and references therein) as well as the outstanding features in
the RM sky \cite{sk80} were all attributed to the disturbed local magnetic
fields. No large-scale magnetic fields in the Galactic halo were
recognized.

\vspace{1mm}
\noindent{\bf D. Fields near the Galactic center} 

Near the Galactic center vertical filaments were observed \cite{ymc84} 
and interpreted as illumination of vertical magnetic fields with mG 
strength \cite{ym87}.

\vspace{1mm}
\noindent{\bf E. Strength of regular and random fields } 

The strength of large-scale regular field is about 2~$\mu$G
\cite{hq94,rk89}, and the total field is about 6$\mu$G. Therefore, random
field is stronger than regular field.  By adopting a single-cell-size
model for the turbulent field, Rand \& Kulkarni obtained a turbulent field
strength of 5~$\mu$G with a cell size of 55 pc \cite{rk89}, and Ohno \&
Shibata got 4 - 6~$\mu$G for the random field with an assumed cell size in
the range 10 - 100 pc \cite{os93}. Extensive discussions on the strength and
energy of the random field and regular field can be found in
\cite{hei96b}. Rand \& Lyne found evidence for stronger fields towards the
Galactic center \cite{rl94}.

\vspace{1mm}
\noindent{\bf F. Other Unknowns}

There was no information about the variation of field strength with
Galactocentric radius or Galactic height, although there were some hints for
such variations \cite{bkb85}. It is understandable that the
fields in the arm region could be more tangled than these in the interarm
regions \cite{hq94,hei96b}, but not much more information was available. There was
no consideration of the spatial energy spectrum, i.e., the magnetic field
strength on different scales, although turbulence in interstellar medium was
known already to follow the Kolmogorov spectrum on small scales.

\section{The Galactic magnetic fields: progress in the last decade}
Compared to the knowledge a decade ago about the above terms, significant
progress has been made on many aspects as we describe below.  Pulsars have
unique advantages as probes of the large-scale Galactic magnetic
field. Their distribution throughout the Galaxy at approximately known
distances allows a true three-dimensional mapping of the large-scale field
structure. Furthermore, combined with the measured DMs, pulsar RMs give us a
direct measure of the mean line-of-sight field strength along the path,
weighted by the local electron density. In last decade, a large number of
pulsars have been discovered by Parkes pulsar surveys \cite{mld+96,mlc+01,mhl+02},
and many of them are distributed over more than half of the Galactic
disk. The RMs of these pulsars provide a unique opportunity for
investigation of the magnetic field structure in the inner Galaxy. Compared
with about 200 pulsars RMs available in 1993 \cite{tml93}, now in total, the
RMs of 550 pulsars have been observed \cite{mhth05}, and about 300 of them
by our group \cite{qmlg95,hmq99,hml+06}.

\vspace{-2mm}
\subsection{The magnetic field structure versus the spiral arms: new consensus} 
A bisymmetric spiral model for magnetic fields in local area ($<$ a few kpc)
has been established by using pulsar RM data \cite{hq94,id99,hmq99}. The
pitch angle of the magnetic fields is about $-8\degr$, with reversals of 
magnetic field directions. The new analysis of starlight polarization data 
\cite{hei96} also gives a pitch angle of large-scale magnetic fields about 
$-8^o$, coincident with that from pulsar data. We therefore conclude that
the large-scale magnetic fields in our Galaxy, at least in the local region,
follow the spiral structure and probably have the same pitch angle of spiral
arms.
\begin{figure*}[htb]
\begin{center}
\includegraphics[angle=270,width=29pc]{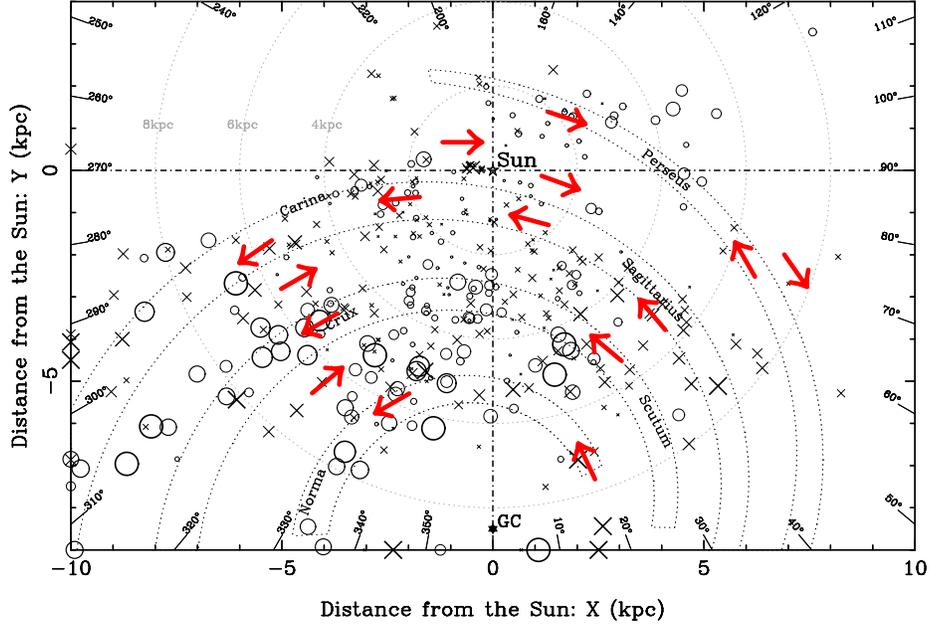}
\vspace{-3mm}
\caption{The RM distribution of 374 pulsars with $|b|<8\degr$, projected
onto the Galactic Plane. The linear sizes of the symbols are proportional to
the square root of the RM values, with limits of 9 and 900 rad m$^{-2}$. The
crosses represent positive RMs, and the open circles represent negative
RMs. The approximate locations of four spiral arms are indicated. The
large-scale structure of magnetic fields derived from pulsar RMs are
indicated by thick arrows. See reference \cite{hml+06} for details.}
\end{center}
\vspace{-3mm}
\end{figure*}

\subsection{Discrimination of Models} 

The limited pulsar RM data and only measurements in local Galactic regions
give room for three models to survive. Discrimination between different
models is complicated by small-scale irregularities of field structures and
sparse measurements. However, the measured pitch angle of the magnetic
fields using different approaches, as we have mentioned above, is hard
evidence to rule out the concentric ring model \cite{rk89,rl94} which has
zero pitch angle. See also detailed analysis by Indrani \& Deshpande
\cite{id99} and discussions by Han et al. \cite{hmq99,hml+06}. The
axisymmetric model of Vall\'ee \cite{val96} suggests that the field near the
Norma arm is clockwise, which has been disapproved by pulsar RM data 
\cite{hmlq02,hml+06}. The large-scale fields in the Galactic disk derived
from new pulsar RM data \cite{hml+06} suggest a tighter BSS field structure.

\begin{figure*}[htb]
\begin{center}
  \begin{minipage}[t]{0.6\linewidth}
\centerline{\psfig{figure=ers.ps,angle=270,width=80mm}}
  \end{minipage}%
  \begin{minipage}[t]{0.4\textwidth}
\centerline{\psfig{figure=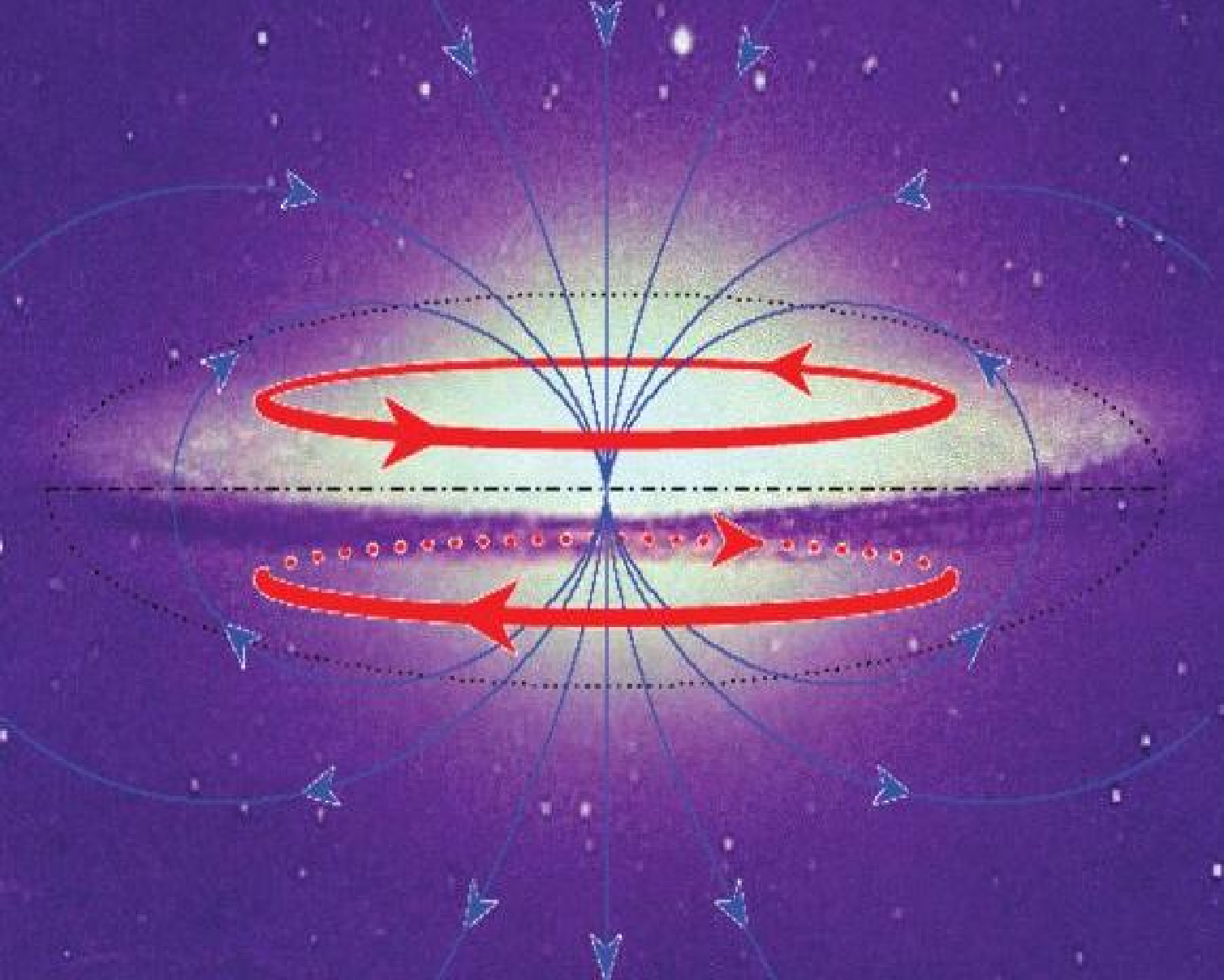,angle=0,width=50mm}}
  \end{minipage}%
\vspace{-4mm}
\caption{The antisymmetric rotation measure sky, derived from RMs of
  extragalactic radio sources after filtering out the
  outliers of anomalous RM values, should correspond to the magnetic field 
  structure in the Galactic halo as illustrated on the right \cite{hmbb97,hmq99}.}
\end{center}
\vspace{-5mm}
\end{figure*}

\subsection{Field structure in the Galactic disk: new measurements}

We observed more than 300 pulsar RMs \cite{qmlg95,hmq99,hml+06}, most of
which lie in the fourth and first Galactic quadrants and are relatively
distant. These new measurements enable us to investigate the structure of
the Galactic magnetic field over a much larger region than was previously
possible. We even detected counter-clockwise magnetic fields in the most
inner arm, the Norma arm \cite{hmlq02}. A more complete analysis
\cite{hml+06} gives such a picture for the coherent large-scale fields
aligned with the spiral-arm structure in the Galactic disk, as shown in
Fig.1: magnetic fields in all inner spiral arms are counterclockwise when
viewed from the North Galactic pole.  On the other hand, at least in the
local region and in the inner Galaxy in the fourth quadrant, there is good
evidence that the fields in interarm regions are similarly coherent, but
clockwise in orientation.  There are at least two or three reversals in the
inner Galaxy, occurring near the boundary of the spiral arms
\cite{hmq99,hml+06}.  The magnetic field in the Perseus arm can not be
determined well, although Brown et al. argued for no reversal \cite{btwm03},
using the negative RMs for distant pulsars and extragalactic sources which
in fact suggest the interarm fields both between the Sagittarius and Perseus
arms and beyond the Perseus arm are predominantly clockwise.

\subsection{Field structure in the Galactic halo}
The magnetic field structure in halos of other galaxies is difficult to
observe. Our Galaxy is a unique case for detailed studies, since polarized
radio sources all over the sky can be used as probes for the magnetic fields
in the Galactic halo.

From the RM distribution in the sky, Han et al. identified the striking
antisymmetry in the inner Galaxy respect to the Galactic coordinates
\cite{hmbb97,hmq99}, as being a result from the azimuth magnetic fields in
the Galactic halo with reversed field directions below and above the
Galactic plane (see Fig.2). Such a field can be naturally produced by an A0
mode of dynamo (see reference \cite{wk93} for a review). The observed
filaments near the Galactic center should result from the dipole field in
this scenario.  The local vertical field component of $\sim$0.2~$\mu$G 
\cite{hq94,hmq99} may be related to the dipole field in the solar vicinity.

I have shown \cite{han04} that the RM amplitudes of extragalactic radio
sources in the mid-latitudes of the inner Galaxy are systematically larger
than those of pulsars, indicating that the antisymmetric magnetic fields are
not local but are extended towards the Galactic center, far beyond the
pulsars. We are observing more RMs of extragalactic radio sources and
modeling the RM sky with a various magnetic field structure in the Galactic
halo. The azimuthal halo fields with reversed directions above and below the
Galactic plane could simply result from a shearing of the dipole fields by
differentially rotating layers of the ISM.

\subsection{Field strength on different scales}

Interstellar magnetic fields exist over a broad range of spatial scales,
from the large Galactic scales to the very small dissipative scales, but
with different field strength. Knowledge of the complete magnetic energy
spectrum can offer a solid observational test for dynamo and other theories
for the origin of Galactic magnetic fields \cite{bk05}.

Estimation of the large-scale field strength \cite{hq94,hml+06} and a
turbulent field strength at a scale of tens of pc \cite{rk89,os93} is only
the first step. It is also possible to get more hints from electron density
fluctuations in interstellar medium, since magnetic fields are also always
frozen in the interstellar gas. The spatial power spectrum of electron
density fluctuations from small scales up to a few pc \cite{ars95} could be
approximated by a single power law with a 3D spectral index $-$3.7, very
close to the Kolmogorov spectrum, which gives us a hint that the magnetic
energy on the small scales to a few pc should have the Kolmogorov spectrum
as well. This was confirmed by Minter \& Spangler who found \cite{ms96} that
structure functions of RM and emission measure were consistent with a
3D-turbulence Kolmogorov spectra of magnetic fields up to 4~pc, but with a
2D turbulence between 4~pc and 80 pc. The RM fluctuations, due to both the
magnetic fields and electron density, are much enhanced in the Galactic
spiral arms than in interarm regions \cite{hgb+06}.

\begin{figure}[htb]
\includegraphics[width=55mm,height=74mm,angle=-90]{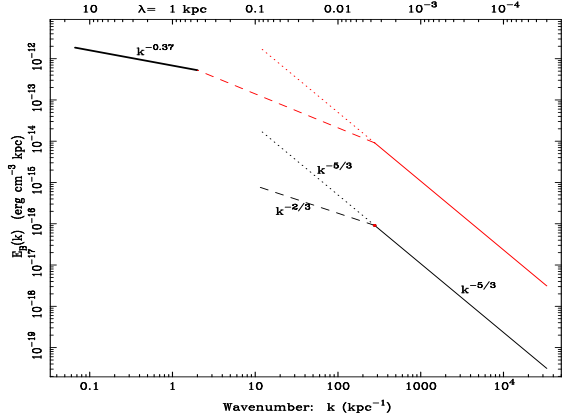}
\vspace{-3mm}
\caption{ Composite magnetic-energy spectrum in our Galaxy.
The large-scale spectrum was derived from pulsar RM data \cite{hfm04}.
The thin solid and dashed/dotted lines at smaller scales are the
Kolmogorov and 2D-turbulence spectra given by Minter \& Spangler 
\cite{ms96}, and the upper one is from new measurements of Minter 
(2004, private email).}
\vspace{-3mm}
\end{figure}

Pulsar RMs are the integration of field strength times electron density over
the path from a pulsar to us. Therefore, RM data of pulsars with different
distances should reflect the fluctuations on different scales. Han et
al. investigated the RM differences at different scales and obtained the
spatial energy spectrum of the Galactic magnetic field in scales between
$0.5<\lambda<15$~kpc \cite{hfm04}, which is a 1D power-law as $E_B(k)\sim
k^{-0.37\pm0.10}$, with $k=1/\lambda$. The rms field strength is
approximately 6 $\mu$G over the relevant scales and the spectrum is much
flatter than the Kolmogorov spectrum for the interstellar electron density
and magnetic energy at scales less than a few pc.

\subsection{Variation of the field strength with Galactocentric radius}
Stronger regular magnetic fields in the Galactic disk towards the Galactic
Center have been suggested in references \cite{sf83,rk89,hei96b}.
Such a radial variation of total field strength has been derived 
from modeling of the Galactic synchrotron emission (E. Berkhuijsen, Fig.14
in \cite{wie05}) and the Galactic $\gamma-$ray background \cite{smr00}. 
Measurements of the regular field strength in solar vicinity give values 
of $1.5\pm0.4\;\mu$G \cite{rk89,hq94,id99}, but near the Norma arm it is 
$4.4\pm0.9\;\mu$G \cite{hmlq02}.
\begin{figure}[htb]
\includegraphics[angle=270,width=74mm]{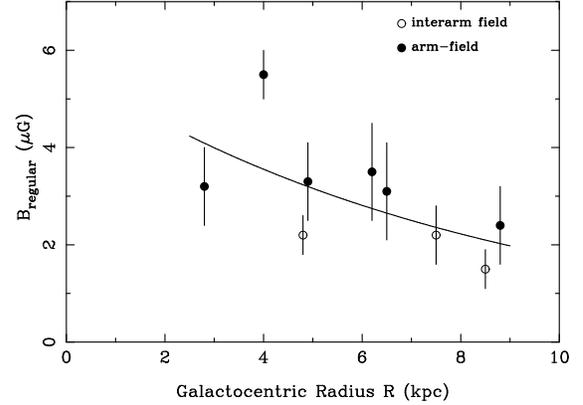}
\vspace{-3mm}
\caption{Variation of the large-scale regular field strength with the
Galactocentric radius derived from pulsar RM and DM data near the 
tangential regions \cite{hml+06}. Note that the ``error-bars'' 
are not caused by the uncertainty of the pulsar RM or DM data, but reflect 
the random magnetic fields in the regions. Filled dots are for arm
regions and small open circles are for interarm regions. The curved line is
a fit of an exponential model.}
\vspace{-3mm}
\end{figure}

With the much more pulsar RM data now available, Han et al. were able to
measure the regular field strength near the tangential points in the 1st and
4th Galactic quadrants \cite{hml+06}, and then plot the dependence of
regular field strength on the Galactoradii (see Fig. 4). Although
uncertainties are large, there are clear tendencies for fields to be
stronger at smaller Galactocentric radii and weaker in interarm regions. To
parameterize the radial variation, an exponential function was used as
following, which not only gives the smallest $\chi^2$ value but also avoids
the singularity at $R=0$ (for $1/R$) and unphysical values at large R (for
the linear gradient). That is,
$
B_{\rm reg}(R) =
B_0 \; \exp \left[ \frac{-(R-R_{\odot})} {R_{\rm B}} \right] ,
$
with the strength of the large-scale or regular field at the Sun,
$B_0=2.1\pm0.3$ $\mu$G and the scale radius $R_{\rm B}=8.5\pm4.7$ kpc.

\subsection{Magnetic fields near the Galactic center}

Progress has been made in two aspects for the region within tens to hundreds
pc of the Galactic Center, both for poloidal fields and for toroidal fields 
\cite{nov05}.

\vspace{1mm}
\noindent{\bf Poloidal fields:} More non-thermal filaments near the Galactic
center have been discovered \cite{lnlk04,nlk+04,yhc04}. The majority of the
brighter non-thermal filaments are perpendicular to the Galactic plane,
indicating a predominantly poloidal fields of $\sim$mG strength, but some
filaments are not, indicating a more complicated field structure than just
the poloidal field. LaRosa et al. detected the diffuse radio emission
\cite{lbs+05} and argued for a weak pervasive field of tens of $\mu$G near
the Galactic Center. The new discovery of an infrared 'double helix' nebula
\cite{mud06} reinforces the conclusion of strong poloidal magnetic fields
merging from the rotated circumnuclear gas disk near the Galactic center.

\vspace{1mm}
\noindent{\bf Toroidal fields:} With the development of polarimetry at mm,
submm or infrared wavelengths, toroidal fields have been observed near the
Galactic center \cite{ncr+03,cdd+03}, complimenting the poloidal fields
shown by the vertical filaments.  Analysis of the much enhanced RMs
\cite{rrs05} of radio sources near the Galactic Center may indicate toroidal
field structure.

\section{The Galactic magnetic fields: work to do}

We have known a lot about the Galactic magnetic fields, but it is far away 
to have a full picture of Galactic magnetic fields. Here I list some
questions which might be answered in next years.

1. Large-scale field structure in the Galactic disk: The RM data of
extragalactic radio sources in the outer Galaxy show no field reversal in
the Perseus arm \cite{btwm03}, while the weak evidence for the reversal
comes from pulsar RMs about $l\sim70^o$ \cite{hmq99,wck+04}. RMs of newly
discovered pulsars from Arecibo pulsar survey \cite{cfl+06} would become
available in or exterior to the Perseus arm, which can settle down this
controversy. The field structure in the 1st Galactic quadrant can be,
but not yet, revealed by more pulsar RM data.

2. Detailed field structure and field strength in the Galactic halo.
We need much more RM data over the all sky. We have observed 1700 RMs
using Effelsberg telescope and will try our best for the model of the
halo field.

3. It is important to know the magnetic energy spectrum from scales 
of 1~pc to 0.5~kpc, which has not been well determined at the moment. The 
strongest field should be the energy-injection-scale, which should be 
a few pc from supernova remnants. We have very little measurements 
of the fields on scales around 10~pc, which is extremely important 
for the discrimination of mechanisms for the maintenance or generation 
of magnetic fields. It is also necessary to determine the spectrum at 
small scales in different parts of the Galactic disk.

4. We have to understand the magnetic field near the Galactic center.
There is consensus on the field structure \cite{yhc04,mud06} but not 
the field strength \cite{lbs+05,ym87}. More data and physical analysis 
are desired to make a coherent picture.

5. The field strength must vary with the Galactic height and the
Galactocentric radius. At present, we haven't had a good measure on
$B\sim B(z)$ yet. We should be able to model it according to currently
available data.

\section{Concluding remarks}

In the last decade, there has been significant progress in studies of 
Galactic magnetic fields, mainly due to the availability of a large number of
newly observed RMs of pulsars. Further pulsar rotation measure observations,
especially for interarm regions and especially in the first Galactic
quadrant, would be especially valuable to confirm the large-scale magnetic
field structure in the Galactic disk. An improved RM database for the
whole sky will enable us to probe details of the magnetic fields in the
Galactic halo. Future detailed modeling of the global magnetic field 
structure of our Galaxy should match all measurements of the fields
in different directions or locations, including the field near the Galactic 
center.

\section*{Acknowledgments}
I am very grateful to colleagues who have collaborated to make the progress
described in this review: Dr. R.N. Manchester from Australia Telescope
National Facility, CSIRO, Prof. G.J. Qiao from Peking University (China),
Prof. Andrew Lyne from Jodrell Bank Observatory (UK), and Dr. Katia
Ferri\'ere from Observatory of Midi-Pyr\'en\'ees (France). 
The author is supported by the National Natural Science Foundation 
of China (10521001 and 10473015).


\label{lastpage}

\end{document}